# THE ARTIFACTS OF COMPONENT-BASED DEVELOPMENT

**M. Rizwan Jameel Qureshi, Shaukat Ali Hayat**
Dept. of Computer Science,
COMSATS Institute of Information Technology
rjamil@ciitlahore.edu.pk, Shaukat.Ali@ciitlahore.edu.pk
Ph # (92-42-5431602) Cell # (03334492203)

*ABSTRACT: Component based development idea was floated in a conference name "Mass Produced Software Components" in 1968 [1]. Since then engineering and scientific libraries are developed to reuse the previously developed functions. This concept is now widely used in SW development as component based development (CBD). Component-based software engineering (CBSE) is used to develop/ assemble software from existing components [2]. Software developed using components is called component ware [3]. This paper presents different architectures of CBD such as ActiveX, common object request broker architecture (CORBA), remote method invocation (RMI) and simple object access protocol (SOAP). The overall objective of this paper is to support the practice of CBD by comparing its advantages and disadvantages. This paper also evaluates object oriented process model to adapt it for CBD.*

## 1. INTRODUCTION

Process model is a framework having standard procedures used in an organization to analyze, design and develop software. The goal of process models is to establish a relationship between quality of process model and quality of software [4]. Quality of software means it must function according to customer requirements. It must possess all quality parameters defined by McCall [3] in 1977. Most of the offered tools help to develop information systems by meeting the requirements of process models such as structured and object oriented [5,6]. Many process models, like Waterfall, Rapid Application development (RAD) [7], object oriented (OO) [8,9], Rational Unified Process (RUP), Component Based Development (CBD) and Agile [3,10], are proposed in the history of software development. There are attempts to adapt and improve component-based development [2,11,12] software development from last few years to cope the needs of software industry.

Reuse of software components concept has been taken from manufacturing industry and civil engineering field. Manufacturing of vehicles from parts and construction of buildings from bricks are the examples.

*Car manufacturers have not been so successful if they have not used standardized parts/components.*

Software companies have used the same concept to develop software in standardized parts/components. Software components are shipped with the libraries available with software. *Microsoft Corporation* and *Sun Microsystems* are two major software-providing organizations. These companies have provided parts/components with their software to market themselves successful and their tools are widely used in software industry. For example, Microsoft Visual Basic (VB), JBuilder, Microsoft.Net, Power Builder, Delphi, etc provides an IDE (Integrated Development Environment). IDE helps programmers to develop programs in such a short time which is only possible using an IDE. As the name indicates, IDE provides an environment in which components are available in the toolbox or in the reference library like a car assembling plant. We do not need to develop the components during the assembling of the car but they are there and we timely assemble them. Similarly in IDE, the standard components such as text box, label box and command button are available in the toolbox and we just integrate and use them. Visual Basic also allows executing a program without compilation in order to just see the results of the coding. We do not need separate programs to develop the components or to run its program. This is the reasons VB is most widely used Rapid Application Development Tool (RAD). Vendors other than Microsoft and Sun Microsystems are also providing components with their tools to make them successful.

Different people have defined the component in different ways. The most popular definition is "*Each reusable binary piece of code is called a component*" [13]. The component concept is similar to object concept of object-oriented programming. A component is an independent part of the system having complete functionalities. A component is designed to solve a particular purpose, such as command button and text box of VB [14]. The component is like a pattern that forces the developers to use the predefined procedures and meets the specifications to plug it into the new SW components.

Section 2 to 5 describes several architectures, proposed up to now, for CBD [3,15]. Section 6 evaluates a process model for CBD. Section 7 and 8 describe main advantages and disadvantages of CBD.

## 2. Microsoft Architecture

Microsoft Corporation ActiveX components have three major categories, based on the type of applications where these are used.

- Component Object Model (COM)
- Distributed Component Object Model (DCOM)
- Object linking and embedding (OLE)

COM was originally developed to create the desktop applications. DCOM is being developed to create remote or distributed applications. OLE is used to integrate one SW tool with another tool seamlessly, e.g., integrating MS Word with MS Excel [16]. COM+ is an extended version of COM. The COM, DCOM and OLE technologies as a whole are called ActiveX technology. Microsoft Corporation has introduced a new set of tools i.e., Microsoft.NET. It is a framework that uses common language runtime (CLR) environment. ActiveX technology can be used with the .NET technology.

By using Microsoft Corporation tools, programmers have facility to develop their own components. Programmers can develop components in any tool/language that supports COM. A component developed in VB can be used with ASP, Visual C, Visual JAVA and Visual C++. COM technology has two main benefits.

- It helps the programmer to integrate tools of different vendors.
- Components developed in different programming languages communicate with each other. This ability of components to intercommunicate is also known as interoperability. Use of Crystal report with VB or ASP is an example of COM. Word application of one vendor integrating Spreadsheet application of another vendor is another example of COM.

### 3. Common Object Request Broker Architecture

Object Management Group (OMG) has introduced a standard for the intercommunication of the components known as **common object request broker architecture (CORBA)**. The CORBA divides the architecture of different programming languages developed by different vendors, e.g. VB procedure can be called to C or C++ and vice versa. The language used for intercommunication among the components is known as interface definition language (IDL). The procedure used to call the remote objects is known as remote procedure call (RPC). The service that is used to communicate one component with another is called object request broker (ORB) [3]. The components used by the CORBA need to be registered in a repository. The architectural design of CORBA is shown in figure 1. The client component request is passed to IDL stub and then to client operating system. The client operating system passes that request to server operating system. The server OS sends that message to server IDL stub and then to the server component. The server component replies by providing the necessary data to IDL stub. The IDL server stub sends that message to the server OS. The server OS then sends that reply to the client OS. The client OS sends the reply to the client IDL stub and then to the client object.

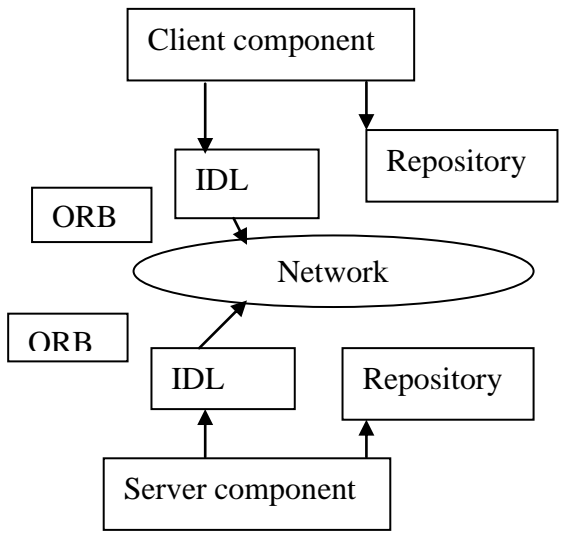

Figure 1 The Architectural Design of CORBA [17]

There are two main limitations of the CORBA architecture: the programmer needs to learn a new programming language, i.e. IDL; CORBA supports a limited number of data types using JAVA.

### 4. Sun Microsystems Architecture

Sun Microsystems architecture allows the programmers to use either enterprise Java bean or RMI technology. Beans perform the same operation as the ActiveX components. RMI is used to call the components of a Java application into another Java application over the network. The method to call the objects is the same as is followed by CORBA. The main limitation of RMI is that it is applicable only for Java applications. The programmer has to use CORBA if he/she wants to call the component of an application developed in a tool other than JAVA [18].

### 5. Simple Object Access Protocol (SOAP)

A group of corporations (Microsoft, IBM and Lotus) have developed SOAP protocol. It is a combination of HTTP and XML. XML is based on the document object model (DOM). Unlike the HTML, that is used only to display the data, XML is used to manipulate the data. It is based on the object-oriented concept. XML document has parent and child objects that are also called as nodes. The procedure used to call the object of one application into another application is the RPC as is used by CORBA [15].

### 6. Process Model of CBD

The object oriented process model is the only process model that indicates the reuse of existing SW parts [8].

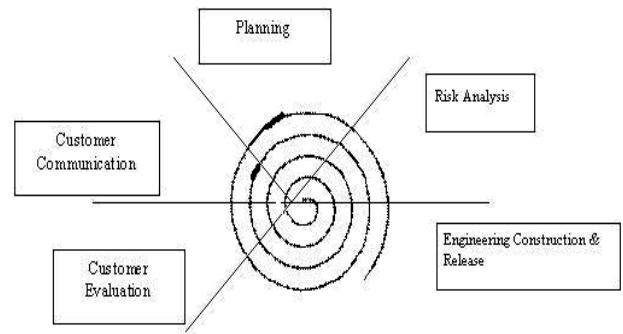

Figure 2 OO Process Model

That model can be modified to implement component-based development. The main phases of process model are shown, in figure 2.
- Customer Communication
- Planning
- Risk Analysis
- Engineering, construction and release
- Customer Evaluation

Customer is communicated to gather the basic requirements like conventional software development. Feasibility assessments are made to prepare a cost-benefit analysis sheet. Risk analysis phase is started if customer approves feasibility assessments. Risk analysis phase of the OO process model can be modified according to the CBD process model. This is the phase where an analyst gathers the detailed requirements of the system to be developed. A domain analysis is performed to find a suitable architecture for the application to be developed [19]. An architectural model of application is developed that enables software engineer to:
- evaluates the efficiency of the design;
- Judge the options of design;
- minimize the potential threats coupled to software development [3].

Software engineering team evaluates the requirements to identify those requirements that can be fulfilled from reusable components by using a repository [20]. Software engineering methods are applied to develop new components for those requirements that cannot be fulfilled from already developed components.

Reusable components need qualification, adaptation and composition. Component qualification makes sure that the selected component will:
- execute the desired functionality;
- integrate easily into the structural design of new application;
- demonstrate the quality attributes (e.g., reliability, performance, usability) required.

The relationships among the components are identified. The properties and behaviors of the components are identified as well [21]. Constraints are always there even after components qualify for reuse into new application, such as technical and integration constraints. Components are wrapped to manage constraints during adaptation. Composition involves integration of components into the architecture of new application. An infrastructure is required to achieve composition. Infrastructure will:
- connect components;
- provide interoperability;
- execute operations;

Infrastructure constitutes a collection of four architectural elements to attain component composition [22].
- Data Exchange Model- Procedures to access and communicate data be written for reusable components e.g., drag and drop, cut and paste and API based data exchange.
- Automation- Kinds of SW tools, subroutines and scripts can be applied to achieve interoperability between components.
- Structured Storage- Technology hold varied data as a single component document to index, search and access components (a single data arrangement rather than set of independent files) e.g., ActiveX and DOM.
- Architectural Models- CORBA, RMI, Microsoft COM/DCOM.

The core objective of this phase is to reuse maximum components, rather than reinventing the wheel.

Engineering, construction & release phase of the OO model matches the requirements of CBD process model. The newly components are designed, developed and tested. The integration and system tests of newly developed as well as of the reused components are performed. IDL is coded to integrate the components if the programmer is using CORBA or RMI architecture. The customer evaluation phase of OO process model fits to the requirements of the CBD process model.

## 7. Advantages of CBD Approach

There are many advantages of developing the applications using the CBD [17].
I. Reusable
II. Interoperable
III. Up-Gradable
IV. Saving the programmers from complexity
V. Time effective
VI. Cost effective
VII. Makes programmers Efficient
VIII. Reliable
IX. Improved Quality

**Reusable:** It is an important advantage of developing the applications. Developing a mail transfer web application system in ASP can be an example application using the CBD. CDONTS component has been provided by Microsoft Corporation to develop e-mail systems. This component can be used in similar applications. The built-in components within a particular tool also reflect the benefits of reusability approach such as Microsoft VB 6.0 text box and command button objects. The programmer needs intensive coding to develop a text box with its complete functionality using the C language whereas a text box with its complete functionality in VB 6.0 is a component and can be used frequently without writing a single line of code. That makes the VB 6.0 the most widely used Rapid Application Development (RAD) tool. Reusability helps us to concentrate on adding more complex functionalities in our applications rather than focusing on developing basic components.

CBD architectures allow the components to intercommunicate with each other and this makes them **interoperable**. The interoperability facilitates the programmers to integrate one application with other applications. The information systems used by ATM machines of different banks are integrated with each other. One link system is the benefit of interoperability provided by CBD.

**Up gradable:** Up gradation of the web applications is easy if some new component has been introduced during the life of application. The programmer has just to replace that previous component with the new component in the application without a change in the front-end code of the application. The client-server architecture is used for the development of distributed applications. There are three types of client-server architecture. Single tier architecture is composed of integrating presentation, application logic and data layers into single SW application such as MS Excel and Access SW. Two-tier architecture integrates presentation and application logic layers into one SW application and data layer is handled by second SW application. This can be explained by taking an example of a business application that is coded by using VB or ASP to handle the presentation and application logic layers where data layer is handled by MS Access or SQL server database. The three tier architecture is also called multiple or n-tier architecture. The presentation, application logic and data layers are handled by three SW applications. The front-end tool such as VB or ASP handles the presentation layer; the COM or COM+ component handles business logic layer while MS Access or SQL server database handles data layer.

These days the programmers prefer to use three-tier architecture to develop business applications. The business logics are continuously changing due to current business environment. Therefore, it is very easy for the programmer to replace the COM component without changing the front-end VB or ASP code and back-end MS Access or SQL server database.

The programmer does not need to understand how any component is working. He should know how to integrate that component with the application. This is similar to a car driver who does not need to know how a Honda Car engine operates. The programmer can concentrate more on main features of the system rather than wasting time on designing the basic components of the application such as text box and command button. CBD saves the programmers from the **complex programming** of designing the development environment. CBD helps to reuse the components again and again in the similar applications that result in **time** and **cost saving**. It also makes the programmers **efficient.**

The reusable components are already thoroughly tested and maintained when these were used for the first time. Therefore the reusable components are **reliable** in terms of usage as compared to the newly developed components. As stated by Wayne C. Lim, the defect rate for reuse code is 0.9 defects per kilo line of code (KLOC) where as it is 4.1 defects per KLOC for new code [23]. He made a comparison between two applications, one was developed with reuse and other was developed without reuse. The improvement in the development time was 51% for the application that was developed from the reusable components.

The **quality** of the application developed from reusable components is also significantly improved. The reusable components are highly tested and maintained. The newly developed components need a lot of testing and maintenance and still there are some bugs left that may be reported when they are integrated with other components. The quality of the SW is affected because of such maintenance. As stated by Henry and Faller [24], the quality of the SW is 35 % improved because of the reusable components.

**8. Disadvantages of CBD Approach**

Following are the main disadvantages of component-based development in terms of reuse:
  I. Customization
 II. Adaptability
III. Integration
 IV. Security
  V. Efficiency

Customization of an already developed component according to the requirements of new application is a major issue in CBD. The developers also face a problem to adapt a component to a new platform if it were not developed for that platform. The integration of a reusable component into new component is also a major problem faced by most of the developers. Security is another major concern for the developers who reuse the components available over the Internet. There may be a virus inside that component and may pass all the information of the business organization to attacker, who uses such an application. Efficiency of the SW applications developed using CBD is also debatable. The component to be reused may have extra functionalities that may be a requirement when it was developed. The new application that does not require extra functionalities becomes less efficient because of the loading time of those functions.

**9. CONCLUSION**

This paper evaluates the main architectures of CBD. The OO process model is also modified for CBD. Main advantages and disadvantages of CBD are also illustrated. Therefore it can be concluded that CBD results in cost effective, saves time and productive for the SW development community.